\documentclass[aps,amsmath,floats,floatfix, twocolumn,
superscriptaddress,nofootinbib,showpacs]{revtex4-1} 

\usepackage[colorlinks, pdfborder={0 0 0}, plainpages=false]{hyperref}
\usepackage{graphicx}
\usepackage{array}
\usepackage{xspace}
\usepackage[usenames,dvipsnames]{color}
\usepackage{amssymb}
\usepackage[normalem]{ulem} 
\usepackage{mathrsfs}

\newcommand{\Caltech}{\affiliation{Theoretical Astrophysics 350-17,
    California Institute of Technology, Pasadena, CA 91125, USA}}
\newcommand{\Cornell}{\affiliation{Center for Astrophysics and Planetary Science, Cornell University, Ithaca, New York 14853, USA}}
\newcommand{\CITA}{\affiliation{Canadian Institute for Theoretical
    Astrophysics, 60 St.~George Street, University of Toronto,
    Toronto, ON M5S 3H8, Canada}} %
\newcommand{\GWPAC}{\affiliation{Gravitational Wave Physics and
    Astronomy Center, California State University Fullerton,
    Fullerton, California 92834, USA}} %
\newcommand{\RIT}{\affiliation{Center for Computational Relativity and Gravitation, School of Mathematical Sciences, Rochester Institute of Technology, 85 Lomb Memorial Drive, Rochester, New York, 14623, USA}}
\newcommand{\JPL}{\affiliation{Caltech JPL, Pasadena, California 91109, USA}}

\begin{document}

\title{Modeling the source of GW150914 with targeted numerical-relativity simulations}

\author{Geoffrey~Lovelace}\GWPAC
\author{Carlos~O. Lousto}\RIT
\author{James~Healy}\RIT
\author{Mark A.~Scheel}\Caltech
\author{Alyssa Garcia}\GWPAC
\author{Richard~O'Shaughnessy}\RIT
\author{Michael Boyle}\Cornell
\author{Manuela Campanelli}\RIT
\author{Daniel A.~Hemberger}\Caltech
\author{Lawrence~E.~Kidder}\Cornell
\author{Harald P.~Pfeiffer}\CITA
\author{B\'{e}la Szil\'{a}gyi}\Caltech\JPL
\author{Saul A.~Teukolsky}\Cornell
\author{Yosef Zlochower}\RIT

\date{\today}

\begin{abstract}
In fall of 2015, the 
two LIGO detectors 
measured the gravitational wave signal GW150914, which
originated
from a pair of merging black holes~\cite{Abbott:2016blz}.
In the final 0.2 seconds (about 8 gravitational-wave cycles)
before the amplitude reached its maximum, 
the observed signal swept up in amplitude and frequency, from 35 Hz to 150 Hz.
  The theoretical gravitational-wave signal for merging
  black holes, as predicted by general
  relativity, can be computed only by full numerical relativity,
  because analytic approximations fail near the time of merger.
  Moreover, the nearly-equal masses, moderate spins, and small number
  of orbits of GW150914 are especially straightforward and efficient to
  simulate with modern numerical-relativity codes.
   In this paper, we report the modeling of GW150914 with 
  numerical-relativity simulations, using 
  black-hole masses and
  spins consistent with
those inferred from LIGO's measurement~\cite{TheLIGOScientific:2016wfe}.
  In particular, we employ two independent
  numerical-relativity codes that use
  completely different analytical and numerical methods to model
   the same   merging black holes and to compute the
   emitted gravitational
     waveform; we find
  excellent agreement between 
      the waveforms produced by the two independent codes.  
    These results 
    demonstrate the validity, impact, and potential of current and future 
    studies using rapid-response, targeted numerical-relativity simulations for
    better understanding gravitational-wave observations.
\end{abstract}


\maketitle

\section{Introduction}\label{sec:intro}

On September 14, 2015, the Advanced Laser Interferometer
  Gravitational-Wave Observatory (Advanced LIGO) directly measured
  gravitational waves  for the first
  time~\cite{Abbott:2016blz}, giving birth to a new era of astronomy.
  The waves were emitted by a pair of black holes with
  masses $36_{-4}^{+5}$\,$M_{\odot}$\ and $29_{-4}^{+4}$\,$M_{\odot}$~\cite{Abbott:2016blz,TheLIGOScientific:2016wfe} that
  orbited each other, collided, and merged about 1.3 billion
  years ago. The 
  gravitational wave signal from this event, named GW150914, is consistent with
  general relativity~\cite{TheLIGOScientific:2016src}.
LIGO has already observed a second gravitational-wave signal
    from merging black holes (called GW151226)~\cite{Abbott:2016nmj} and a third   possible signal (called LVT151012)~\cite{TheLIGOScientific:2016pea}; 
  many more such observations are expected soon.
  Extrapolating from the observations of GW150914 and GW151226
  and including LVT151012, Advanced LIGO is
  expected to observe between roughly 5  to tens of
  gravitational-wave signals from merging black holes during its next six-month
  observing run (O2) starting in
  2016~\cite{TheLIGOScientific:2016pea}.

 The GW150914 observation 
included 8 gravitational-wave cycles, covering the late
  inspiral, merger, and ringdown phases of the binary (cf. Fig.~2 of
  Ref.~\cite{Abbott:2016blz});  
this late phase of a binary-black-hole (BBH) merger
can be described accurately only by directly solving the full
equations of general relativity.
To extract and validate robust conclusions about the astrophysical and fundamental significance of these events \cite{Abbott:2016blz,TheLIGOScientific:2016wfe,GW150914-ASTRO,TheLIGOScientific:2016src}, 
correct and complete solutions to Einstein's equations will be critical, and can be obtained only through direct
numerical simulation.

The first attempts to solve the general relativity field equations numerically
date to the 1960s, when Hahn and Lindquist~\cite{HahnLindquist1964} 
performed the first studies.
Smarr  followed these early efforts
with some success in the 1970s~\cite{SmarrThesis,1977NYASA.302..569S}.  
  The field matured in the 1990s, when a large collaboration
  of  research groups 
  worked together toward solving
  the ``Grand Challenge'' of evolving 
 BBHs~\cite{Abrahams1998}.  
The crucial final breakthrough
  came in 2005,
when three groups \cite{Pretorius2005a, Campanelli2006a,Baker2006a}
 devised two completely independent techniques
to  numerically solve the  BBH
 problem. The first
solution \cite{Pretorius2005a} handled the spacetime singularity by 
excising  the regions interior to the black hole horizons, while the second technique
\cite{Campanelli2006a,Baker2006a}, dubbed the ``moving punctures approach,'' 
used singularity avoiding slices of the black hole spacetimes. 

Since then, through considerable effort by many research groups
  (reviewed, e.g., in Refs~\cite{Choptuik15,sperhake2014numerical,Tiec:2014lba,Hannam:2013pra,Pfeiffer:2012pc}), each of these techniques have matured,
  yielding  two distinct, 
independent approaches to modeling
   BBHs with numerical relativity. 
Important analytic and numerical developments 
\cite{Loffler:2011ay,Mroue:2013PRL,Szilagyi:2015rwa}
have improved the accuracy
and  efficiency
of the codes implementing
 each technique.
 Each
 technique has enabled numerical relativists to begin constructing
 catalogs of 
 BBHs and associated gravitational waveforms~\cite{Ajith:2012az,Hinder:2013oqa,Mroue:2013PRL,Pekowsky:2013ska,Healy:2014yta,Clark:2014fva,Husa:2015iqa}. A
BBH can be
  characterized by seven intrinsic parameters: the ratio
  $q$ of the holes' masses and the spin-angular-momentum vectors
  $S_1^i$ and $S_2^i$ of each hole; as a result, these catalogs must include
  many  BBH
  simulations
  to span the parameter space of 
  BBH events that LIGO could observe.
  Note that early work has already used
    numerical-relativity waveforms
    for detection and parameter estimation in LIGO, for the injection
    of waveforms~\cite{Aylott:2009ya,Aylott:2009tn,Ajith:2012az,Aasi:2014tra},
into LIGO noise, 
    and for establishing common error measures 
    to establish standard of required waveform accuracy \cite{Hinder:2013oqa}.

In this paper, we present a detailed comparison of the targeted
numerical BBH simulations that modeled GW150914 in Figs. 1--2 of 
Ref.~\cite{Abbott:2016blz}, provided by two codes: SpEC and LazEv. We
chose the parameters (masses and spins) of these simulations to be
consistent with estimates of the
parameters for
GW150914~\cite{Abbott:2016blz,TheLIGOScientific:2016wfe}. Note that we are not presenting any additional
  information on parameter estimates in this paper. The
  parameters we chose are consistent with LIGO's observation of GW150914, but we
  could have chosen different parameters that are also consistent with
  the observation. As discussed in
  Ref.~\cite{Abbott:2016blz,TheLIGOScientific:2016wfe}, there is
  considerable uncertainty in parameter estimates for GW150914,
  particularly for the black-hole spins.

By
comparing the results from these two codes, our study extends the
statement made in the caption of Fig.~1 of Ref.~\cite{Abbott:2016blz}:
that the numerical relativity calculations shown there are ``confirmed
to 99.9\%.''
Our investigation extends previous
validation studies of each code
internally~\cite{Baker-Campanelli-etal:2007,Hannam:2009hh}, and
against one another \cite{2012CQGra..29l4001A}, using common standards
for waveform accuracy \cite{Hinder:2013oqa}, to current
versions of both
numerical-relativity codes for the important case of modeling GW150914. 

SpEC and LazEv are completely independent.
They use different formulations of Einstein's
equations, so they assume different decompositions of $G_{\mu\nu}=8\pi
T_{\mu\nu}$ into evolution equations and constraints, and they solve for
different dynamical variables.  They use different methods to choose
coordinates and different methods of handling the spacetime
singularities inside the black holes.  They use different analytic and
numerical methods for generating constraint-satisfying initial data,
and they use different geometries for the numerical grid.  They use
different numerical methods for the spacetime evolution, for refining
the numerical grid, and for extracting the gravitational waveforms
from the evolved variables.   They share no subroutines in common.  But
despite these vast differences, we show in this paper that the two
codes produce the same physics. This is a very strong test of both
codes and of the analytical and numerical methods underlying them.

Moreover, our comparison itself began
independently. While we based both simulations on the same mass and
spin parameter estimates, we only began discussing the
comparison when lower resolutions had
finished and higher resolutions were already under way.
We made no special effort to tune the
codes or the simulations to agree with each other.
In this way, we
have demonstrated the agreement of our codes under realistic working
conditions, where multiple groups independently perform rapid follow up to
a LIGO observation.

These results build confidence that both numerical
methods produce consistent physics, which in turn builds confidence in 
current and future studies making use of targeted numerical-relativity
simulations to follow up LIGO observations.
  Targeted numerical-relativity simulations already feature in recent studies
  directly comparing data for
  GW150914 to catalogs of numerical-relativity
  simulations~\cite{Abbott:2016apu,Jani:2016wkt} and in recent studies 
  injecting numerical-relativity waveforms into LIGO data for
  GW150914,
  to help assess systematic errors in approximate waveform
  models~\cite{TheLIGOScientific:2016wfe, Abbott:2016izl,
    TheLIGOScientific:2016src}.

This paper is organized as follows.
In Sec.~\ref{sec:SXS} we briefly describe the formalism and implementation of
SpEC,  used by the Simulating eXtreme Spacetimes (SXS) collaboration
to numerically evolve 
BBHs. In Sec.~\ref{sec:RIT} we describe the
different formalism and code implementation of LazEv, used by the
Rochester Institute of Technology (RIT) group. 
Table \ref{tab:methods} summarizes the independent methods these two codes employ to construct and evolve initial data
for black holes.  
In  Sec.~\ref{sec:results}, we directly   compare the waveforms produced by each approach to one another.
In Sec.~\ref{sec:discussion} we conclude with
a discussion of the significance of those comparisons and
implications for future comparisons of observations with numerical-relativity calculations.

\begin{table*}
\begin{tabular}{|>{\raggedright}p{2.2in}||>{\raggedright}p{2.2in}|>{\raggedright\arraybackslash}p{2.2in}|}
\hline 
&  \textbf{LazEv} & \textbf{SpEC} \\ \hline \hline
\multicolumn{3}{|l|}{\hspace*{2cm} \bf \emph{Initial data}}\\ \hline
Formulation of Einstein constraint equations & conformal method using Bowen-York solutions~\cite{Murchadha-York:1974b,Pfeiffer2003b,Bowen-York:1980}& conformal thin sandwich \cite{York1999,Pfeiffer2003b}  \\ \hline
Singularity treatment & puncture data~\cite{Brandt1997} &  quasi-equilibrium black-hole excision~\cite{Caudill-etal:2006,Cook2004,Cook2002} \\ \hline
Numerical method & pseudo-spectral~\cite{AnsorgBruegmann2004} & pseudo-spectral~\cite{Pfeiffer2003} \\ \hline
Achieving low orbital eccentricity & post-Newtonian inspiral \cite{Husa-Hannam-etal:2007}
 & iterative eccentricity removal~\cite{Pfeiffer-Brown-etal:2007,PhysRevD.83.104034} \\ \hline \hline
\multicolumn{3}{|l|}{\hspace*{2cm}\bf \emph{Evolution}} \\ \hline
Formulation of Einstein evolution equations & BSSNOK~\cite{NOK87,Shibata1995,Baumgarte99}
 & first-order generalized harmonic with constraint damping~\cite{Lindblom:2007,Friedrich1985,Pretorius2005c,Pretorius2005a} \\ \hline
Gauge conditions & evolved lapse and shift~\cite{Bona1997,Alcubierre2002,vanMeter2006} & damped harmonic~\cite{Szilagyi:2009qz} \\ \hline
Singularity treatment & moving punctures~\cite{Campanelli2006a,Baker2006a} & excision \cite{Scheel2006} \\ \hline
Outer boundary treatment & Sommerfeld & minimally-reflective, constraint-preserving \cite{Lindblom:2007,Rinne2007} \\ \hline
Discretization & high-order finite-differences~\cite{Husa2007,Zlochower:2005bj} & 
pseudo-spectral methods \\   \hline
Mesh refinement & adaptive mesh refinement~\cite{Schnetter2003b} & 
domain decomposition with spectral adaptive mesh 
refinement~\cite{Pfeiffer2003,Szilagyi:2009qz}  \\ \hline
  \end{tabular}
\caption{A comparison of the two  independent 
    numerical relativity codes
  described in the text. 
  Each  code uses different
  techniques to construct and evolve initial data for
   BBHs and to extract the emitted gravitational radiation. This table is based on Table I of
  Ref.~\cite{Pfeiffer:2012pc}.  \label{tab:methods}
}
\end{table*}

\section{Simulations using pseudospectral, excision methods\label{sec:SXS}}

Simulations labeled SXS are carried out using the Spectral Einstein
Code (SpEC)~\cite{SpECwebsite}.  Given initial BBH parameters, a
corresponding weighted superposition of two boosted, spinning
Kerr-Schild black holes~\cite{Lovelace2008} is constructed, and then
the constraints are solved~\cite{York1999,Pfeiffer2003,Ossokine:2015yla} by a
pseudospectral method to yield
quasi-equilibrium~\cite{Caudill-etal:2006,Lovelace2008} initial data.
Small adjustments in the initial orbital trajectory are made
iteratively to produce initial data with low
eccentricity~\cite{Pfeiffer-Brown-etal:2007,Buonanno:2010yk,Mroue:2012kv}.

The initial data are evolved using a first-order
representation~\cite{Lindblom2006} of a generalized harmonic
formulation~\cite{Friedrich1985, Garfinkle2002, Pretorius2005c} of
Einstein's equations, and using damped harmonic
gauge~\cite{Lindblom2009c,Choptuik:2009ww,Szilagyi:2009qz}.  The
equations are solved pseudospectrally on an
adaptively-refined~\cite{Lovelace:2010ne,Szilagyi:2014fna} spatial grid that
extends from pure-outflow excision boundaries just inside apparent
horizons~\cite{Scheel2009,Szilagyi:2009qz,Hemberger:2012jz,Ossokine:2013zga,Scheel2014}
to an artificial outer boundary.  
Adaptive time-stepping automatically achieves
 time steps of approximately the Courant limit.
On the Cal State Fullerton cluster, ORCA, the simulation
achieved a typical evolution speed of $O(100 M)/{\rm day}$ for the highest
resolution (here we measure simulation
time in units of $M$, the total mass of the
binary).
After the holes merge, all variables
are automatically interpolated onto a new grid with a single excision
boundary inside the common apparent
horizon~\cite{Scheel2009,Hemberger:2012jz}, and the evolution is
continued.  Constraint-preserving boundary
conditions~\cite{Lindblom2006, Rinne2006, Rinne2007} are imposed on
the outer boundary, and no boundary conditions are required or imposed
on the excision boundaries.

We use a pseudospectral fast-flow algorithm~\cite{Gundlach1998} 
to find apparent horizons, and we compute spins on these apparent horizons
using the approximate Killing vector formalism of Cook, Whiting, and
Owen~\cite{Cook2007, OwenThesis}.

Gravitational wave extraction is done by three independent methods:
direct extraction of the Newman-Penrose quantity $\Psi_4$ at finite
radius~\cite{Scheel2009,Boyle2007,Pfeiffer-Brown-etal:2007},
extraction of the strain $h$ by matching to solutions of the
Regge-Wheeler-Zerilli-Moncrief equations at finite
radius~\cite{Buchman2007,Rinne2008b}, and Cauchy-Characteristic
Extraction~\cite{Bishop:1997ik,Winicour2005,Gomez:2007cj,Reisswig:2012ka,Handmer:2014}.
The latter method directly provides gravitational waveforms at future
null infinity, while for the former two methods the waveforms are
computed at a series of finite radii and then extrapolated to
infinity~\cite{Boyle-Mroue:2008}.  Differences between the different
methods, and differences in extrapolation algorithms, can be used to
place error bounds on waveform
extraction. These waveform extraction errors are important for
the overall error budget of the simulations, and are typically on the
order of, or slightly larger than, the numerical truncation
error~\cite{Taylor:2013zia,Chu:2015kft}. In this paper, the waveforms
we compare use Regge-Wheeler-Zerilli-Moncrief extraction and
extrapolation to infinity. We have verified that our choice of extrapolation
order does not significantly affect our results. We have also checked
that corrections to the wave modes~\cite{Boyle2015a} to account for
a small drift in the coordinates of the center of mass have a negligible
effect on our results.

\section{Simulations using finite-difference, moving-puncture methods}\label{sec:RIT}

RIT simulations evolve BBH data sets using the {\sc
LazEv}~\cite{Zlochower:2005bj} implementation of the moving puncture
approach~\cite{Campanelli2006a,Baker2006a} with the conformal
function $W=\sqrt{\chi}=\exp(-2\phi)$ suggested by
Ref.~\cite{Marronetti:2007wz}.  For the simulation presented here, we use
centered, sixth-order finite differencing in
space~\cite{Lousto:2007rj},  a fourth-order Runge-Kutta time
integrator,\footnote{Note that we do not upwind the advection terms.}
and a 7th-order Kreiss-Oliger dissipation operator.
This sixth-order spatial finite difference scheme
allows us to gain
a factor $\sim4/3$ in efficiency with the respect to the eighth-order implementation, because 
it reduces the number of ghost zones from 4 to 3.
We also allowed for a Courant factor CFL$=1/3$ instead of the
previous CFL$=0.25$ \cite{Zlochower:2012fk} 
gaining another speedup factor of 4/3. 
We verified that for this relaxing of the time integration step
we still conserve the horizon masses and spins of the individual
black holes during evolution and the phase of the gravitational
waveforms to acceptable levels. This plus the use of the new
XSEDE supercomputer {\it Comet} at 
SDSC~\footnote{https://portal.xsede.org/sdsc-comet} lead to typical
evolution speeds of $250M/{\rm day}$ on 16 nodes for $N100$ and similar
for the higher resolution runs given the good weak scaling of LazEv. 
Note that our previous \cite{Lousto:2013oza,Lousto:2015uwa}
comparable simulations averaged $\sim100M/{\rm day}$.

RIT's code uses the {\sc EinsteinToolkit}~\cite{Loffler:2011ay,
einsteintoolkit} / {\sc Cactus}~\cite{CACTUS} /
{\sc Carpet}~\cite{Schnetter2003b}
infrastructure.  The {\sc
Carpet} mesh refinement driver provides a
``moving boxes'' style of mesh refinement. In this approach, refined
grids of fixed size are arranged about the coordinate centers of both
holes.  The {\sc Carpet} code then moves these fine grids about the
computational domain by following the trajectories of the two black holes.

We use {\sc AHFinderDirect}~\cite{Thornburg2004} to locate
apparent horizons.  We measure the magnitude of the horizon spin using
the {\it isolated horizon} (IH) algorithm detailed in
Ref.~\cite{Dreyer2003} and as implemented in Ref.~\cite{Campanelli:2006fy}.
Note that once we have the horizon spin, we can calculate the horizon
mass via the Christodoulou formula
\begin{equation}
{m} = \sqrt{m_{\rm irr}^2 + S^2/(4 m_{\rm irr}^2)} \,,
\end{equation}
where $m_{\rm irr} = \sqrt{A/(16 \pi)}$, $A$ is the surface area of
the apparent horizon, and $S$ is the spin angular momentum of the black hole (in
units of $M^2$).  In the tables below, we use the variation in the
measured horizon irreducible mass and spin during the simulation as a
measure of the error in computing these quantities.  
We measure radiated energy,
linear momentum, and angular momentum, in terms of the radiative Weyl
Scalar $\Psi_4$, using the formulas provided in
Refs.~\cite{Campanelli:1998jv,Lousto:2007mh}. However, rather than
using the full $\Psi_4$, we decompose it into $\ell$ and $m$ modes and
solve for the radiated linear momentum, dropping terms with $\ell >
6$.  The formulas in Refs.~\cite{Campanelli:1998jv,Lousto:2007mh} are
valid at $r=\infty$.  We extract the radiated energy-momentum at
finite radius and extrapolate to $r=\infty$. We find that the new
perturbative extrapolation described in Ref.~\cite{Nakano:2015pta} provides the
most accurate waveforms. While the difference of fitting both linear and
quadratic extrapolations provides an independent measure of the error.

\section{Results}\label{sec:results}

\begin{figure*}
\includegraphics[angle=0,width=1.96\columnwidth]{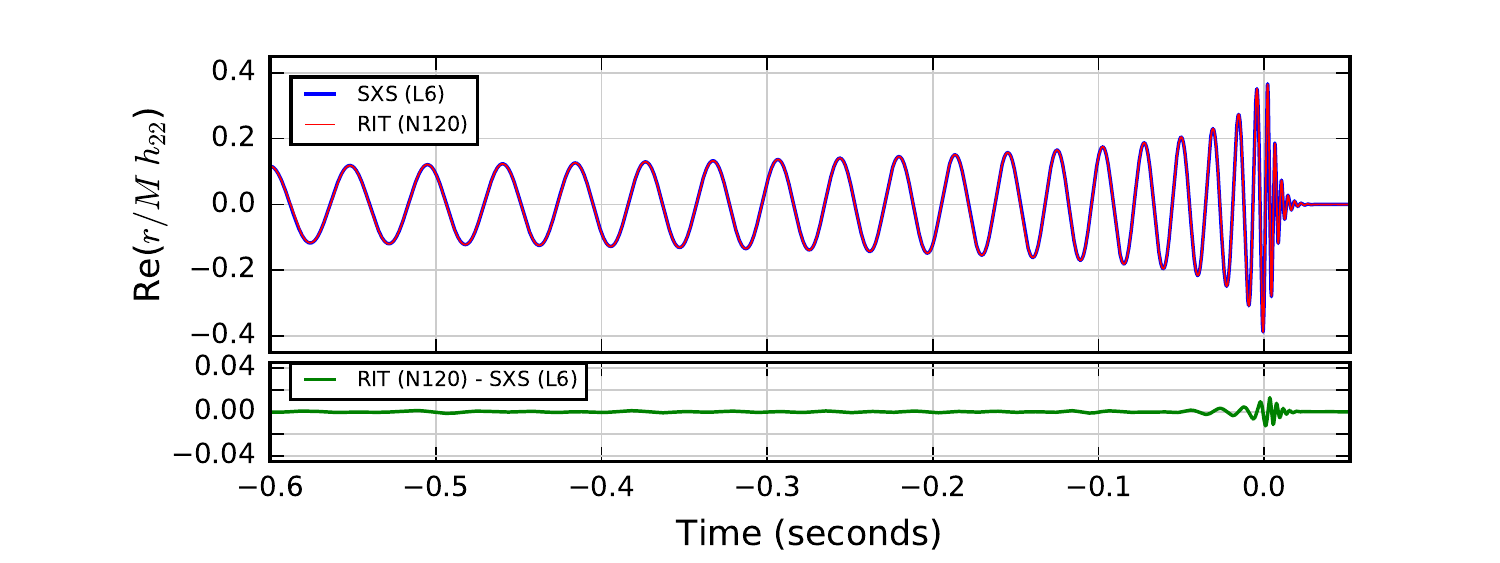}
\caption{A comparison of the $(\ell,m)=(2,2)$ mode extracted from the two fiducial simulations from SpEC and LazEv. Time is shown in seconds for a total mass of $70 M_{\odot}$. The bottom panel zooms in to show the difference between the SpEC and LazEv (2,2) modes, using
  resolutions are $N120$ (LazEv) and $L6$ (SpEC). Here, we take
  a conservative approach to assessing differences in the waveforms: we apply a
  constant time shift to each waveform, so that the peak (2,2) amplitude is
  at $t=0$, and a constant phase offset to each waveform, so that
  the phase is zero radians at $t=-0.6\mbox{ s}$.
  Table~\ref{tab:match} and Fig.~\ref{fig:comparison3}, in contrast, assess
  differences by computing the match, a comparison weighted by LIGO's noise.
Note that lower resolution versions of these waveforms were 
used in the comparison in the caption of Fig.~1 of
Ref.~\cite{Abbott:2016blz}, which reported
the agreement (match) of these waveforms as 99.9\%, using LIGO's 
noise curves for GW150914~\cite{Abbott:2016blz}.
\label{fig:comparison1}}
\end{figure*}

To model GW150914, we used each of the two codes to simulate
a nonprecessing, unequal mass binary with the spin of the larger
black hole aligned with the orbital angular momentum,
and the spin of the smaller black hole antialigned.
The initial data parameters for both codes are given in
Tables~\ref{tab:ID}--\ref{tab:dimlessID}\footnote{Note that data from
  the SXS configuration, including the gravitational waveform and the masses
  and spins as functions of time, are publicly available as simulation
  SXS:BBH:0305 at \url{http://www.black-holes.org/waveforms}}. 
These tables show that the parameters of the two simulations are not
exactly identical, primarily because we constructed and evolved the
initial data independently but also because making the parameters
exactly the same is challenging, given how different the methods for
specifying initial data are in the two approaches.  The largest difference
is that the SXS simulation starts earlier in the inspiral than the RIT
case, but also the two simulations have different orbital
eccentricities and very small differences in the initial masses.

Both  simulations start with a relatively small
 binary separation, so 
the entire evolution through coalescence requires a time of roughly a few thousand
$M$, where $M$ is the sum of the Christodoulou masses of both
black holes.
Because the mass ratio is near unity and the spins are small, this simulation
is not difficult to perform for modern numerical relativity codes;
for both codes considered here, this simulation requires
about 7 to 10 days to complete.
For each code, both simulations were repeated several times using
different values of a parameter controlling the numerical resolution.
Increasing the resolution results in higher accuracy, but requires
more computation time and resources.  Running multiple resolutions provides checks
that the results converge with increasing resolution, and also provides
an error estimate.  The LazEv simulation was performed at three resolutions
labeled $N100$, $N110$, and $N120$, where $N110$ and $N120$ represent a
global increase
of the finite difference resolution by factors of 1.1 and 1.2, 
respectively, compared to the $N100$ case.  The SpEC simulation
was performed at multiple resolutions 
labeled $L0$ through $L6$ in order of increasing resolution; $L6$ 
represents a spectral adaptive-mesh-refinement error tolerance
that is a factor of $e$ smaller than that of $L5$.

\begin{table*}
\begin{ruledtabular}
\begin{tabular}{lccccccccccccc}
Config. & $x_1/M$ & $x_2/M$ & $P_t/M$  & $P_r/M$   & $m^p_1/M$ & $m^p_2/M$ & $S_1/M^2$ & $S_2/M^2$ & $m_1/M$ & $m_2/M$ & $M_{\rm ADM}/M$\\ 
\hline
RIT & -6.7308 & 5.5192  & 0.083116 & -0.000490 & 0.40207   & 0.51363   & -0.08932  & 0.09963   & 0.45055   & 0.54945   & 0.99141         \\
SXS & -7.8597 & 6.4004 & - & - & - & - & -0.08918 & 0.09975 & 0.45020 & 0.54980 & 0.99235\\
\end{tabular}
\end{ruledtabular}
\caption{Initial data parameters for the quasi-circular
configurations with an spinning smaller mass black hole (labeled 1),
and a larger mass spinning black hole (labeled 2). 
For the RIT configuration,
the punctures are located
at $\vec r_1 = (x_1,0,0)$ and $\vec r_2 = (x_2,0,0)$, with momenta
$P=\pm (P_r, P_t,0)$, spins $\vec S_i = (0, 0, S_i)$, mass parameters
$m^p/M$, horizon (Christodoulou) masses $m_1/M$ and $m_2/M$, 
total ADM mass
$M_{\rm ADM}$, and spin angular momenta $S_1/M^2$ and
$S_2/M^2$. 
For the SXS configuration, the apparent horizons begin with centers, in
the asymptotically inertial coordinate frame, at coordinates $(x_1,y_2,0)$ and
$(x_2,y_2,0)$, where $y_1=y_2=-0.0335 M$. 
In both cases, $M=m_1+m_2$ is the sum of the Christodoulou masses of
the two black holes.
The two codes specify initial data differently, so not all parameters
are relevant for both codes.}\label{tab:ID}
\end{table*}

\begin{table*}
  \begin{ruledtabular}
  \begin{tabular}{lccccccc}
    Config. & $q$ & $\chi_1$ & $\chi_2$ & $M\Omega_0$ & $M\dot{a}_0 \times 10^4$ & $d_0/M$ & $e$\\
    \hline
    RIT & 1.220 & -0.4400 & 0.3300 & 0.02118 & -1.1712 & 12.2500 & 0.0012\\
    SXS & 1.221 & -0.4400 & 0.3300 & 0.01696 & -0.5306 & 14.2601 & 0.0008
  \end{tabular}
  \end{ruledtabular}
  \caption{Physical properties of the initial black holes at time $t=0$.
    The table shows the mass ratio $q=m_2/m_1$ and dimensionless
    spins $\chi_1 = S_1/m_1^2$ and $\chi_2 = S_2/m_2^2$. The
    table also shows the initial orbital properties 
  of both configurations. The holes begin at
  coordinate separation $d_0$ with initial orbital angular frequency
  $\Omega_0$ and initial radial expansion $\dot{a}_0$. Each hole's
  initial coordinate
  radial velocity $v_r$ and coordinate
  distance from the center of mass $r_0$
  are related to the expansion by $\dot{a}_0 = v_r/r_0$. These initial data
  result in an initial orbital eccentricity $e$. 
  Here $M=m_1 + m_2$ is the sum of the Christodoulou masses
    of the two black holes.
    \label{tab:dimlessID}}
\end{table*}

The top panel of Fig.~\ref{fig:comparison1} displays the $(\ell,m)=(2,2)$ spin-weighted spherical
harmonic mode of the gravitational waveform
extracted from the two simulations, at the lowest resolution.   
The differences between
these two simulations are not visible at this scale.  
Because of the finite signal-to-noise ratio of
GW150914, the statistical error in the waveform reconstruction reported in Figs.~1--2 of Ref.~\cite{Abbott:2016blz} is far
larger than the differences seen here. Note that we take a deliberately
conservative approach to alignment here: we only apply a constant time shift
(setting the peaks of the $(2,2)$ modes to $t=0$) and a constant phase shift
(setting the phase to zero radians at $t=-0.6\mbox{ s}$. Later
(Fig.~\ref{fig:match} and Table~\ref{tab:match}), we will compare
the mismatch of the waveforms, which (separately for each mode)
optimizes over constant time and phase
offsets and weights the difference inversely to LIGO's noise
[Eq.~(\ref{eq:match})].

\begin{figure}
  \includegraphics[width=0.98\columnwidth]{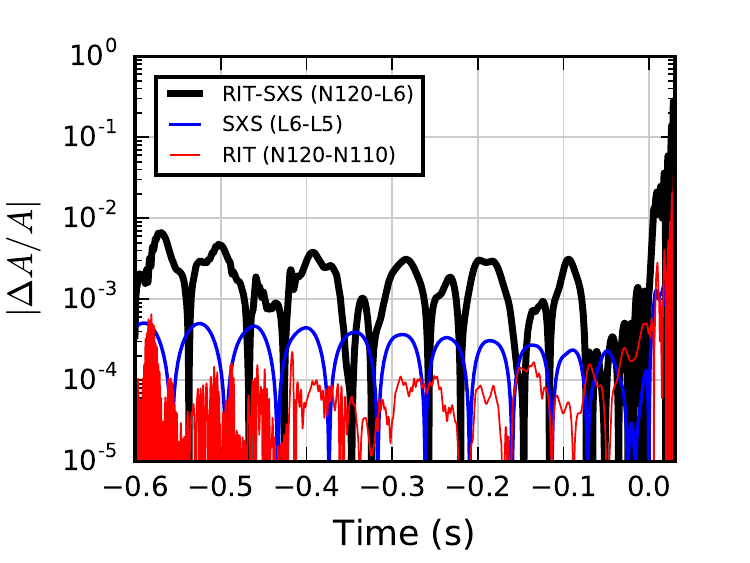}
  \includegraphics[width=0.98\columnwidth]{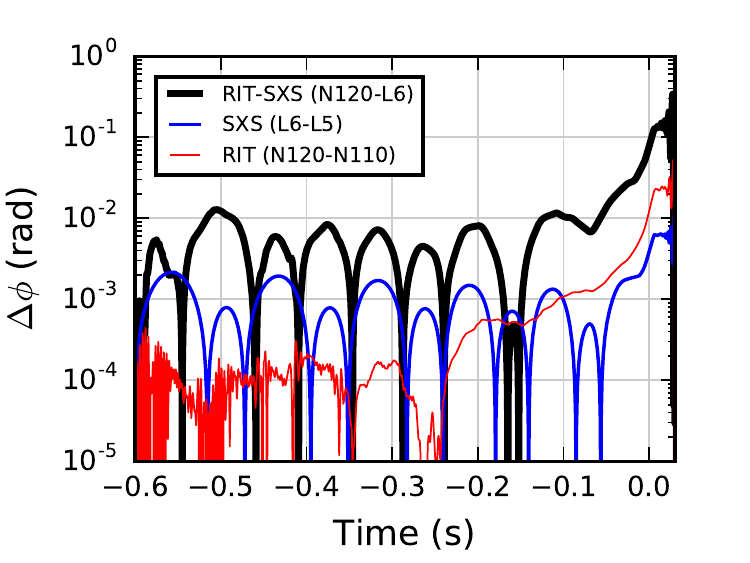}
  \caption{Fractional differences in amplitude $\Delta A/A$ and differences in phase $\Delta \phi$ for the SXS and RIT simulations as a function of
    time (shown in seconds for a total mass of $70 M_\odot$) assuming
    zero inclination. We show 
    numerical error estimates for the SXS and RIT simulations (obtained by subtracting different
    resolutions) and the difference between the highest resolution SXS and RIT simulations. Here, as in Fig. 1, we apply only an overall constant time shift
    and an overall constant phase shift to the waveforms, setting time
    $t=0$ at the peak of the $(\ell,m)=(2,2)$ modes and the phase to
    zero radians at $t=-0.6\mbox{ s}$.
    The differences include both the dominant $(\ell,m)=(2,\pm 2)$ mode and
    also the modes $(2,\pm 1)$,
    $(3,\pm 3)$, $(3,\pm 2)$, $(4,\pm 4)$, and $(4,\pm 3)$.
    \label{fig:comparison3}}
\end{figure}

To better see the differences between the waveforms, the lower
panel of Fig.~\ref{fig:comparison1} zooms in on the difference between
the $(\ell,m)=(2,2)$ modes, and 
Fig.~\ref{fig:comparison3} plots fractional amplitude differences and phase differences, including
not only the $(\ell,m)=(2,2)$ mode but several of the most significant higher
modes. In Fig.~\ref{fig:comparison3}, as in Fig.~\ref{fig:comparison1},
we apply a constant time shift so that the peak $(\ell,m)=(2,2)$ occurs
at time $t=0$ and a constant phase shift so each wave has a phase of zero
radians at time $t=-0.6$ seconds. Differences
between resolutions of each simulation estimate the numerical error.
The differences between the SXS and RIT simulations' highest resolutions is far too small to be visible in a plot like the top panel of Fig.~\ref{fig:comparison1} or
in Fig.~1 of Ref.~\cite{Abbott:2016blz}, which compares the waveforms to LIGO's measured gravitational-wave strain for GW150914.
Nevertheless, the differences are larger than
our estimates of the numerical errors.

  We suspect that the level of disagreement is determined in part
  by effects from small differences in the initial configurations (mass, spin, eccentricity, etc.). Another potential
  source of systematic error is differences in wave extraction: LazEv computes $\Psi_4$ and then integrates
  to find the gravitational-wave strain $h$, while SpEC computed the waves shown in Fig.~\ref{fig:comparison3} by
  matching to Regge-Wheeler-Zerilli-Moncrief solutions. Differences between SpEC and LazEv also become large at late times in the ringdown, when the wave amplitude 
  becomes smaller than the numerical error in the simulations.

  To quantify the magnitude of the differences between these waveforms, we use the match,
\begin{eqnarray}
\mathscr{M} \equiv \frac{\left<h_1\left|\right.h_2\right>}{\sqrt{\left<h_1\left|\right.h_1\right>\left<h_2\left|\right.h_2\right>}},
\end{eqnarray}
as implemented via a complex
overlap as described in Eq.~(2) in Ref.~\cite{Cho:2012ed}:
\begin{eqnarray}
\left< h_1 \left|\right. h_2 \right> & = & 2 \int_{-\infty}^{\infty} \frac{df}{S_n(f)}\left[\tilde{h}_1(f) \tilde{h}_2(f)^* \right],\label{eq:match}
\end{eqnarray} where $\tilde{h}(f)$ is the Fourier transform of $h(t)$ and $S_n(f)$ is the power spectral density of the detector noise (here, taken to be
the advanced LIGO design power spectrum~\cite{LIGO-aLIGODesign-NoiseCurves}). When we compute the match using
Eq.~(\ref{eq:match}), we maximize (separately for each mode)
over an overall constant time shift and
an overall constant phase shift.

For our match calculations, we conservatively adopt the 
design-sensitivity noise power spectrum of advanced LIGO \cite{LIGO-2013-WhitePaper-CoordinatedEMObserving}; 
if we instead would compute matches using the observed O1 sensitivity, the matches would be larger. For instance, the $(\ell,m)=(2,2)$ mode would have
a match of $99.9\%$ (the basis of the ``confirmed to 99.9\%'' statement
in Ref.~\cite{Abbott:2016blz}), instead of
$99.8\%$, as shown here.
Table \ref{tab:match} shows the match between the $(l,m)$ spin-weighted
spherical harmonic gravitational waveform modes of different RIT resolutions versus the SXS L6 resolution, for a total mass of $70 M_\odot$. To account for finite simulation duration, we set the lower frequency of
our match calculation to 
$m \times 11 \mathrm{Hz}$. Note that for the $m=4$ and $m=5$ modes, which are
much less significant than the dominant $(\ell,m)=(2,2)$ mode,
this is above the frequency of $35\mbox{ Hz}$ where the GW150914 signal first entered the LIGO band.
For each mode, the beginning of each waveform (a duration of several $M$ in
time) is tapered, to reduce transient effects in
their Fourier transforms. 
For the dominant (2,2) mode, the match is close to unity for all RIT resolutions.  
  Most other significant modes also display a high degree of agreement,
  particularly for $\ell \le 3$. The table also shows the overlap of L6 with
  itself (with minimum frequency $22\mbox{ Hz}$ in all cases, to indicate the relative importance of each mode. Some modes that
  are much less significant than the $(2,2)$ mode have low matches but are
  (except for some of the least significant modes) convergent,
  suggesting that higher
  numerical resolution is necessary to accurately compute these high order
  modes. 

\begin{figure}
\includegraphics[angle=0,width=0.98\columnwidth]{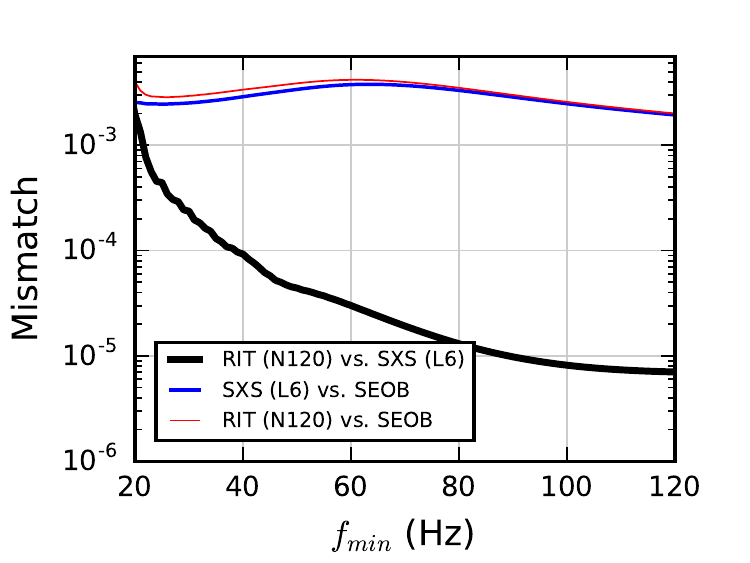}
\caption{Match between SXS and RIT full numerical (2,2)-mode waveforms, against one another and against a corresponding
  SEOBNRv2 template, for our
  fiducial parameters, scaled to a total detector frame binary system mass of $70M_\odot$.  For a sense of scale, at this mass the SXS
and 
RIT waveforms start at $15.7 \rm Hz$ and $19.5 \rm Hz$, respectively.  
The SEOB waveform is evaluated starting at $10 \rm Hz$. The match calculations
adopt the Advanced LIGO design noise spectrum~\cite{LIGO-2013-WhitePaper-CoordinatedEMObserving}.
\label{fig:comparisonEOB}\label{fig:match}
}
\end{figure}

To provide a sense of scale for this comparison, we also evaluate the match between the two simulations' $(\ell,m)=(2,2)$ modes and
the same modes of a template waveform with the same parameters (i.e.,
the same mass ratio and spins) generated by
SEOBNRv2~\cite{Taracchini:2013rva}, one of the approximate, analytic waveform models
used to infer GW150914's properties in Ref.~\cite{TheLIGOScientific:2016wfe}.
Figure~\ref{fig:comparisonEOB} shows the mismatch ($1$ minus the match), 
for successively higher starting cutoff frequencies, $f_{min}$, above 20 Hz.  
At high frequencies, the two simulations are much 
more consistent with one another than with the semianalytic SEOBNR model.
Below about $20 \rm Hz$, 
the two simulations disagree with each other and with 
SEOBNRv2, but this
is because the RIT simulation starts at $19.5 \rm Hz$ and the SXS simulation
starts at $15.7 \rm Hz$, and because the waveforms from both simulations are
tapered.
Note that in order to respond quickly to GW150914, 
to reduce computation time both of these simulations were chosen 
to start at higher frequencies than typical for full numerical simulations.
Therefore, the low-frequency mismatch shown in Fig.~\ref{fig:comparisonEOB}
overstates the differences between the two codes for longer, more typical
simulations.

\begin{table}
\begin{ruledtabular}
\begin{tabular}{cccccr}
$\ell$ & $m$ & $ N100 $ & $ N110 $ & $ N120 $ & $\left<h_{\ell m}^{L6}|h_{\ell m}^{L6}\right>$ \\
\hline
2&	0&	0.8854&	0.8863&	0.8870& 9.82\\
2&	1&	0.9905&	0.9914&	0.9908& 16.78\\
2&	2&	0.9980&	0.9980&	0.9980& 927.74\\
\hline
3&	0&	0.7822&	0.8146&	0.8356& 1.02\\
3&	1&	0.9517&	0.9569&	0.9582& 1.52\\
3&	2&	0.9978&	0.9980&	0.9981& 28.59\\
3&	3&	0.9927&	0.9933&	0.9933& 42.17\\
\hline
4&	0&	0.3603&	0.3581&	0.3554& 0.05\\
4&	1&	0.7910&	0.8348&	0.8616& 0.17\\
4&	2&	0.9074&	0.9425&	0.9562& 1.79\\
4&	3&	0.9844&	0.9909&	0.9938& 2.50\\
4&	4&	0.9863&	0.9886&	0.9901& 40.95\\
\hline
5&	0&	0.3638&	0.4050&	0.4458& 0.01\\
5&	1&	0.2994&	0.3652&	0.4227& 0.01\\
5&	2&	0.6108&	0.6176&	0.6392& 0.14\\
5&	3&	0.7813&	0.8709&	0.9197& 0.32\\
5&	4&	0.9705&	0.9815&	0.9879& 2.49\\
5&	5&	0.9315&	0.9552&	0.9696& 4.94\\
\hline
\end{tabular}
\end{ruledtabular}
\caption{Match between individual spherical harmonic modes $(\ell,m)$
  of the SXS and RIT waveforms, using the advanced LIGO design
  sensitivity.
The successively higher resolution simulations from RIT, labeled as
$N100,N110,N120$ are compared to the L6 (highest) resolution run from SXS. The minimal frequency is taken as $f_{min}=11m\mbox{ Hz}$ for $m>1$ and $f_{min}=22 Hz$ for $m=0,1$ for a fiducial
total mass of $M=70M_\odot$. The column labeled $\left<h_{\ell m}^{L6}|h_{\ell m}^{L6}\right>$ shows the overlap of L6 with itself, with a minimum frequency
of $22\mbox{ Hz}$ in all cases, to indicate the significance of
the mode. 
}
\label{tab:match}
\end{table}

The evolution of the aligned spinning binary leads to remnant
masses, spins and recoil velocities as shown in Table~\ref{tab:srem}.
They display an excellent agreement between the two 
codes,
to at least three significant figures,
and they appear convergent 
with increasing resolution. The fraction $E_{\rm rad} / M$
of the initial mass $M$ radiated as
gravitational waves 
can be inferred (via energy conservation) to be 
\begin{equation}
\frac{E_{\rm rad}}{M} = 1 - \frac{m_{\rm rem}}{M}.
\end{equation}
The SXS and RIT simulations agree that $1-m_{rem}/M = 4.80\%$ of
the initial mass is radiated as gravitational waves. 
For GW150914, whose
initial mass in the source frame is
$65_{-4}^{+5} M_{\odot}$~\cite{TheLIGOScientific:2016wfe}, 
the radiated energy predicted by the SXS and RIT simulations, $3.1 M_{\odot}$,
is consistent with the estimate of $3.0\pm0.5 M_{\odot}$ given in
Ref.~\cite{Abbott:2016blz}.

\begin{table}
\begin{ruledtabular}
\begin{tabular}{cccc}
$\#$ & $m_{rem} / M $ & $\chi_{rem}^z$ & $V_{rem}^{xy} (km/s)$ \\  
\hline
N100 & 0.952015 & 0.691961 & 131.79 \\
N110 & 0.952020 & 0.691965 & 133.35 \\
N120 & 0.952021 & 0.691969 & 134.38 \\
L0 & 0.951760 & 0.691863 & 136.78 \\
L2 & 0.951971 & 0.692030 & 137.22 \\
L4 & 0.952000 & 0.692119 & 136.00 \\
L6 & 0.952033 & 0.692085 & 134.17 
\end{tabular}
\end{ruledtabular}
  \caption{Remnant results for spinning binaries. 
    We show the remnant mass $m_{rem}$ in units of the total initial mass $M \equiv m_1+m_2$,
    the remnant dimensionless spin $\chi_{rem}^z \equiv J_{rem}^z / m_{rem}^2 $, and the
    remnant velocity in the x-y plane $V_{rem}^{xy}$. We show results for different LazEv
    resolutions (N100, N110, and N120) and different SpEC resolutions (L0, L2, L4, and L6).
}
\label{tab:srem}
\end{table}

\section{Discussion}\label{sec:discussion}
We have demonstrated that two completely independent codes to evolve binary black holes (SpEC and LazEv) produce very
similar results.   
As shown in 
Fig.~\ref{fig:comparison1} and Table \ref{tab:match}, we find good agreement even with moderately low resolution simulations
(i.e. N100 and L5).  
A detailed convergence analysis, like that summarized
in Table \ref{tab:match}, suggests that both the generalized
harmonic \cite{Pretorius2005a} 
and moving puncture \cite{Campanelli2006a,Baker2006a} 
approaches lead to accurate solutions of the
general relativity field equations.
Given that the initial configurations are not exactly the same
(different eccentricities, slightly different masses and spins), we 
consider this general agreement 
an excellent verification of the analytic formulations, numerical
methods, and code implementations used in both SpEC and LazEv.

The next steps in further verifying the results of numerical relativity
codes will be to consider binary systems with precession and to consider
simulations that follow a larger number of binary orbits.  For these
more demanding tests, it will be more important to start different
codes with closely coordinated initial parameters. This study will
be the subject of a future publication.

Future work also includes
considering cases with more extreme parameters. Here, the simulations'
very good agreement with the SEOB waveform is not surprising, since
the moderate spins and almost equal masses make this an especially easy
region of the parameter space to model. But for higher mass ratios and
more extreme spins, numerical relativity might disagree more strongly
with semianalytic, approximate waveforms,
especially in regions where the semianalytic models have not been
tuned to numerical relativity. Recent studies have begun exploring the
agreement of numerical relativity and approximate, analytic waveforms
in different regions of the BBH parameter
space~\cite{Kumar:2016dhh,Jani:2016wkt,Husa:2015iqa,Kumar:2015tha,Szilagyi:2015rwa}.

However, from the results of Fig.~\ref{fig:comparisonEOB} 
we can already conclude that even if analytic
waveform
models provide a very good approximation to the
true prediction of general relativity,  
full numerical solutions
of Einstein's equations can be visibly more accurate than analytic models.
Targeted followup with numerical relativity is thus an important tool for 
comparing gravitational-wave observation and theory and for
 reliably measuring potential deviations from Einstein's
 theory of gravitation 
\cite{TheLIGOScientific:2016src}.

Our study suggests that both groups' standard production simulations are sufficiently accurate and
  efficient to respond rapidly and comprehensively to further events like GW150914, informing the analysis and
  interpretation of LIGO data.
  Followup simulations of events like GW150914 can be performed on a timescale of days to weeks
  (depending on resolution) and at low computational cost, with confidence that both
  methods produce consistent physics.
This is important for the construction of numerically generated waveform
data banks with simulations from heterogeneous codes and formalisms.

However, numerical-relativity 
  simulations can be considerably
  more costly and challenging elsewhere in the BBH parameter space,
  particularly if they remain in LIGO's band for more orbits,
  such as GW151226,
  or if they have more extreme parameters. For instance,
  the SpEC simulation modeling GW151226 that
  appears in Fig.~5 of Ref.~\cite{Abbott:2016nmj} (SXS:BBH:0317 at
  \url{http://black-holes.org/waveforms}) required approximately
  2 months to complete, and a recent simulation
  similar to those used here to model GW150914 but with
  spins $\chi=+0.96$ for the larger black hole and $\chi=-0.9$ for the smaller
  black hole (SXS:BBH:0306 at
  \url{http://black-holes.org/waveforms}) required approximately two months
  to complete. Future work includes enabling more rapid, targeted follow up
  numerical-relativity simulations for these more challenging cases.

\begin{acknowledgments}
  We are pleased to acknowledge Joshua R.~Smith and Jocelyn S.~Read
  for helpful discussions of this work and Patricia Schmidt for helpful
  discussions and for helping us to implement
  the numerical-relativity injection HDF5 file format~\cite{injectionFormat}
  she developed with Ian
  Harry, 
  which facilitated some of the comparisons in this paper.
 We gratefully acknowledge the NSF for financial support from grant
 Nos.~PHY-0969855, 
 PHY-1212426, PHY-1305730, PHY-1306125, PHY-1307489, PHY-1229173, PHY-1404569,
PHY-1606522, PHY-1607520, ACI-1550436,  
AST-1028087, AST-1333129, AST-1333520, OCI-0832606, and
DRL-1136221. The authors are also
grateful for support from the Sherman Fairchild 
Foundation, the NSERC of Canada, the Canada Research Chairs Program,
and the Canadian Institute for Advanced Research.
Computational resources were provided by
the ORCA cluster at California State University, Fullerton (CSUF), supported by
CSUF, NSF grant No. PHY-142987,
and the Research Corporation for Science Advancement; 
by
XSEDE allocations
TG-PHY060027N and TG-PHY990007N; by NewHorizons and BlueSky Clusters 
at Rochester Institute of Technology, which were supported
by NSF grant Nos. PHY-0722703, DMS-0820923, AST-1028087, and PHY-1229173;
by the GPC supercomputer at the SciNet HPC Consortium~\cite{Scinet}, which
is funded by the Canada Foundation for Innovation (CFI) under the auspices of
Compute Canada, the Government of Ontario, the Ontario Research Fund (ORF)---Research Excellence, and the University of Toronto; 
and by the Zwicky cluster at Caltech, which is supported by the Sherman 
Fairchild Foundation and by NSF award PHY-0960291.
\end{acknowledgments}

\bibliography{References/References,macros/GW150914_refs}

\end{document}